%% file: cooling.tex
\documentclass[prl,twocolumn,floatix,amssymb,amsmath,superscriptaddress,letterpaper]{revtex4}
\usepackage{hyperref}
\usepackage{graphicx}
\usepackage{color}
\usepackage{multibbl}
\usepackage{amsmath}

\newcommand{\ket}[1]{| #1 \rangle}
\def\cO{\mathcal O}

\begin{document}
\title{Demon-like Algorithmic Quantum Cooling and its Realization with Quantum Optics}

\author{Jin-Shi Xu}
\thanks{These authors contributed equally to this work.}
\affiliation{Key Laboratory of Quantum Information, University of Science and Technology of China, CAS, Hefei, 230026, People's Republic of China}

\author{Man-Hong Yung}
\thanks{These authors contributed equally to this work.}
\affiliation{Department of Chemistry and Chemical Biology, Harvard University, Cambridge MA, 02138, USA}

\author{Xiao-Ye Xu}
\affiliation{Key Laboratory of Quantum Information, University of Science and Technology of China, CAS, Hefei, 230026, People's Republic of China}

\author{Sergio Boixo}
\affiliation{Department of Chemistry and Chemical Biology, Harvard University, Cambridge MA, 02138, USA}

\author{Zheng-Wei~Zhou}
\affiliation{Key Laboratory of Quantum Information, University of Science and Technology of China, CAS, Hefei, 230026, People's Republic of China}

\author{Chuan-Feng~Li}
\email{cfli@ustc.edu.cn}
\affiliation{Key Laboratory of Quantum Information, University of Science and Technology of China, CAS, Hefei, 230026, People's Republic of China}

\author{Al\'{a}n Aspuru-Guzik}
\email{aspuru@chemistry.harvard.edu}
\affiliation{Department of Chemistry and Chemical Biology, Harvard University, Cambridge MA, 02138, USA}

\author{Guang-Can Guo}
\affiliation{Key Laboratory of Quantum Information, University of Science and Technology of China, CAS, Hefei, 230026, People's Republic of China}

\date{\today }

\maketitle
{\bf The simulation of low-temperature properties of many-body systems remains one of the major challenges in theoretical~\cite{Jordan2012,Casanova2012,Kassal2008} and experimental~\cite{baugh_experimental_2005,Barreiro2011,Zhang2012} quantum information science. We present, and demonstrate experimentally, a universal cooling method which is applicable to any physical system that can be simulated by a quantum computer~\cite{Lloyd1996,buluta_quantum_2009,Kassal_Simulating_2011}. This method allows us to distill and eliminate hot components of quantum states, i.e., a quantum Maxwell's demon~\cite{toyabe_experimental_2010}. The experimental implementation is realized with a quantum-optical network, and the results are in full agreement with theoretical predictions (with fidelity higher than 0.978). These results open a new path for simulating low-temperature properties of physical and chemical systems that are intractable with classical methods.
}

From quantum field theories~\cite{Jordan2012} to high-Tc superconductivity~\cite{Casanova2012} and chemical reactions~\cite{Kassal2008}, quantum simulation~\cite{Lloyd1996,buluta_quantum_2009,Kassal_Simulating_2011} provides a unique opportunity for both
theorists and experimentalists to explore a domain of science that goes beyond the applicability of any known classical computing method. Modern low-temperature physics has advanced primarily due to the development of efficient cooling methods~\cite{osheroff_superfluidity_2002, Wieman1999}.  The same is also true for quantum information science.  Physical cooling can drive quantum states towards the physical ground states, which are pure states. However, in the context of quantum simulation, pure states are not necessarily ``cooler" than mixed states. For example, pure states with an equal superposition of all eigenstates correspond to infinite-temperature states~\cite{Poulin2009, yung_quantum-quantum_2010}. In order to achieve cooling for quantum simulation, in general, one not only needs to employ physical cooling to avoid decoherence~\cite{Takahashi2011}, but also be able to prepare states that have low entropy in the basis of the eigenstates of the Hamiltonian of the system being simulated.

An attractive approach for achieving cooling for spins (or qubits) is the heat-bath algorithmic cooling (HBAC) method~\cite{Boykin2002}. The main idea of HBAC is to reduce the entropy of spin systems by unevenly distributing more entropy to one of the spins that can release the excess entropy to a heat bath through thermalization. The feasibility of HBAC has been demonstrated experimentally with NMR technology~\cite{baugh_experimental_2005}. However, HBAC is not a universal method for cooling quantum many-body systems; it is primarily employed for preparing polarized spins as initial states, or ancilla qubits, for quantum computation. Furthermore, the temperature of the environment imposes a fundamental limit for the cooling performance~\cite{Schulman2005}.

\begin{figure}[t]
\begin{center}
\includegraphics [width= 1 \columnwidth]{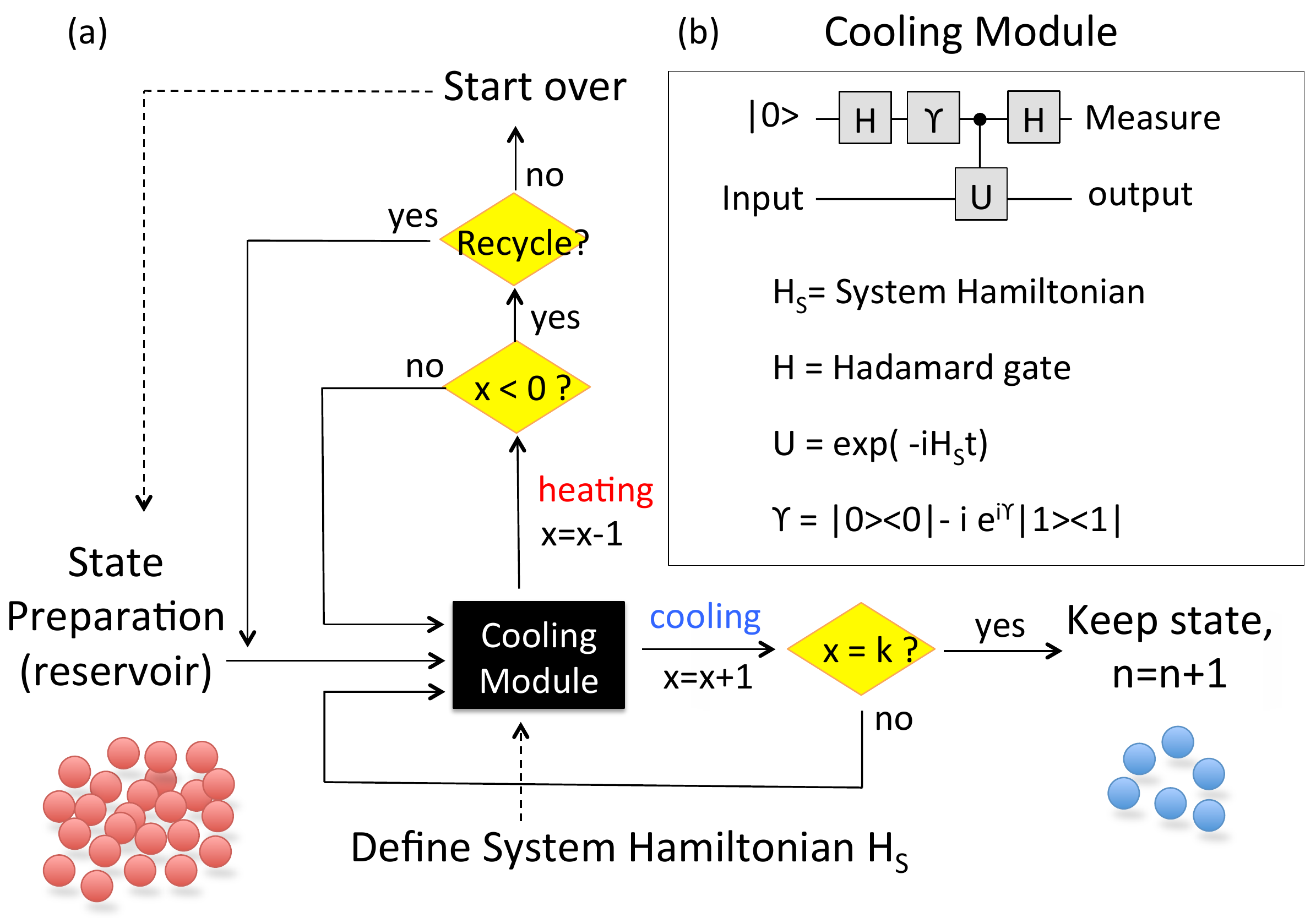}
\end{center}
\caption{(Color online). Overview of the cooling method. (a) Logic diagram of the feedback cooling system. The cooling module produces two outcomes, correlated with heating and cooling. The measurement result can be mapped into the position $x$ of a 1D random walker; when the walker goes beyond its starting position $x=0$ to the negative position $x=-1$,  the particle is either rejected or recycled. (b) The quantum circuit diagram of the cooling module. It includes a controlled evolution for time $t$ and an energy bias parameter~$\gamma$ for optimizing the performance. 
}
\label{fig:logic}
\end{figure}

A more general approach~\cite{Kraus2008, Verstraete2009} for quantum cooling is to engineer a dissipative open-system dynamics to drive quantum states to the ground states of the simulated systems. This dissipative open-system (DOS) approach relaxes two criteria in the standard circuit model~\cite{DiVincenzo2000} for quantum information processing, namely specific initial-state preparation and full-unitary gate decomposition. As the underlying dynamics is based on Lindblad master equations, this approach is shown to be robust~\cite{Weimer2010} against simulation errors. A proof-of-principle demonstration has been experimentally performed using trapped ions~\cite{Barreiro2011}. Despite the advantages the DOS approach can provide, the range of application of it is limited; practically, it is restricted to a class of system involving the so-called frustration-free Hamiltonians~\cite{Verstraete2009}, where the ground state should simultaneously minimize all local terms in the Hamiltonian. This condition is not satisfied by the majority of physical Hamiltonians.


\begin{figure}[t]
\begin{center}
\includegraphics [width= 0.9 \columnwidth]{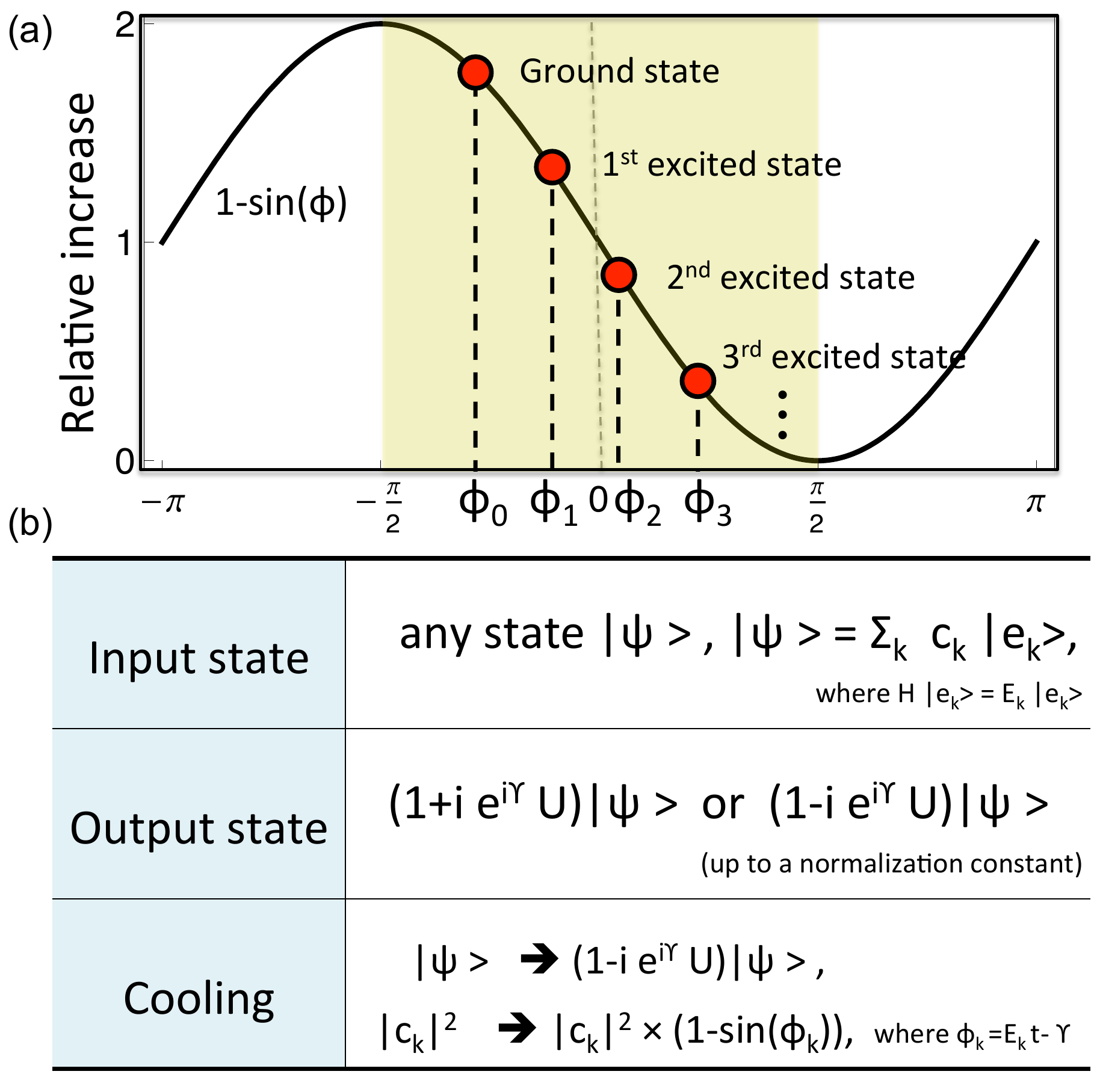}
\end{center}
\caption{(Color online). Basic principle of the cooling method. (a) The relative change in the population of the output state post-conditioned with a ``cooling'' measurement result depends on the eigenenergy~$E_k$: the lower the energy, the higher the gain. Details are summarized in (b). (see also Eq.~\ref{key} and method summary). 
}
\label{fig:cooling_method}
\end{figure}

To overcome the challenges, in this work, we present a universal cooling method applicable to cooling any physical system that can be simulated by a quantum computer. Our method shares some features in the HBAC and the DOS approach, but does not suffer from the limitations of them. Essentially, we replace the requirement of the existence of a heat bath in HBAC by an ensemble of quantum states; thermalization is replaced by swapping with one of the states in the ensemble. As evident by our numerical analysis (see~\cite{SI}), our method does not suffer from similar restrictions in HBAC for cooling spin-1/2 particles~\cite{Schulman2005}. Furthermore, we adopted a feedback approach to result in an effective quantum operation for achieving cooling for quantum simulation. In DOS, each of the quantum jump operators encodes only partial information of the Hamiltonian to be simulated~\cite{Kraus2008, Verstraete2009}; in our case, the jump operators contain full information about the Hamiltonian. This property makes our cooling method universally applicable to any physical Hamiltonian.

To be more specific, for any given Hamiltonian~$H_s$ and any input state, pure state $\left| \psi_{in}  \right\rangle$ or mixed state $\rho_{in}$, the method guarantees that the energy $E \equiv Tr\left( {{H_s}{\rho _{out}}} \right)$ of output state $\rho_{out}$ is less than (or equal to) that of the input state. Importantly, compared with DOS, our method does not increase the qubit resource required for cooling; other than the system qubits, it requires only one extra ancilla qubit, which is the same as in the DOS method. This not only significantly reduces computational resources but also makes the method more robust against noise and errors~\cite{Weimer2010}. Therefore, our cooling method is as realizable as DOS with currently available technologies; we experimentally performed a proof-of-principle demonstration with quantum-optics, which is  reported below.


The core component of the method is the `cooling module' formed by a simple quantum circuit shown in
Fig.~\ref{fig:logic}. It consists of an ancilla qubit initialized into the $\left| 0 \right\rangle$ state, a pair of Hadamard gates $ {\sf H} = \left( {1/\sqrt 2 } \right)\left[ {\left( {\left| 0 \right\rangle  + \left| 1 \right\rangle } \right)\left\langle 0 \right| + \left( {\left| 0 \right\rangle  - \left| 1 \right\rangle } \right)\left\langle 1 \right|} \right]$, a local phase gate ${\sf \gamma}  = \left| 0 \right\rangle \left\langle 0 \right| - i{e^{i\gamma }}\left| 1 \right\rangle \left\langle 1 \right|$, and a controlled unitary operation $U = \exp \left( { - i{H_s}t} \right)$, where $H_s$ is the Hamiltonian to be simulated. We note that the unitary time evolution for most of the physical Hamiltonians can be efficiently simulated with a quantum computer~\cite{Lloyd1996}. For any input state $\left| \psi_{in}  \right\rangle$, the quantum circuit produces the following state~\cite{SI}:
\begin{equation}\label{key}
\Lambda _- \left| {\psi _{in} } \right\rangle \left| 0 \right\rangle  + \Lambda _+ \left| {\psi _{in} } \right\rangle \left| 1 \right\rangle \quad,
\end{equation}
where ${\Lambda _ \pm } \equiv (I \pm i{e^{i\gamma }}U)/2$. Next, the ancilla qubit is measured in the computational basis $\left\{ {\left| 0 \right\rangle ,\left| 1 \right\rangle } \right\}$.  Similar to the working mechanism of DOS~\cite{Kraus2008, Verstraete2009, Weimer2010,Barreiro2011}, the non-unitary operators~${\Lambda _ \pm }$ play similar roles as quantum jump operators for describing the state evolution. Moreover, we found that (see~\cite{SI}) the energy of the resulting state ${\Lambda _ - }\left| \psi_{in}  \right\rangle $ (${\Lambda _ + }\left| \psi_{in}  \right\rangle $), after nomalization of the vector norms, is lower (higher) than that of the input state $\left| \psi_{in}  \right\rangle $. Consequently, the cooling module probabilistically projects any input state $\left| \psi_{in}  \right\rangle $ (or $\rho_{in}$ for mixed states), into either a higher-energy state ${\Lambda _ + }\left| \psi_{in}  \right\rangle $ (ancilla output ${\left| 1 \right\rangle }$) or a lower-energy state ${\Lambda _ - }\left| \psi_{in}  \right\rangle $  (ancilla output ${\left| 0 \right\rangle }$), with respect to the Hamiltonian~$H_s$ being simulated (see Fig.~\ref{fig:cooling_method}a and \ref{fig:cooling_method}b, and~\cite{SI} for a step-by-step derivation). Eq.~\ref{key} represents the key element in our cooling method. 

\begin{figure*}[t]
\begin{center}
\includegraphics [width= 1.8 \columnwidth]{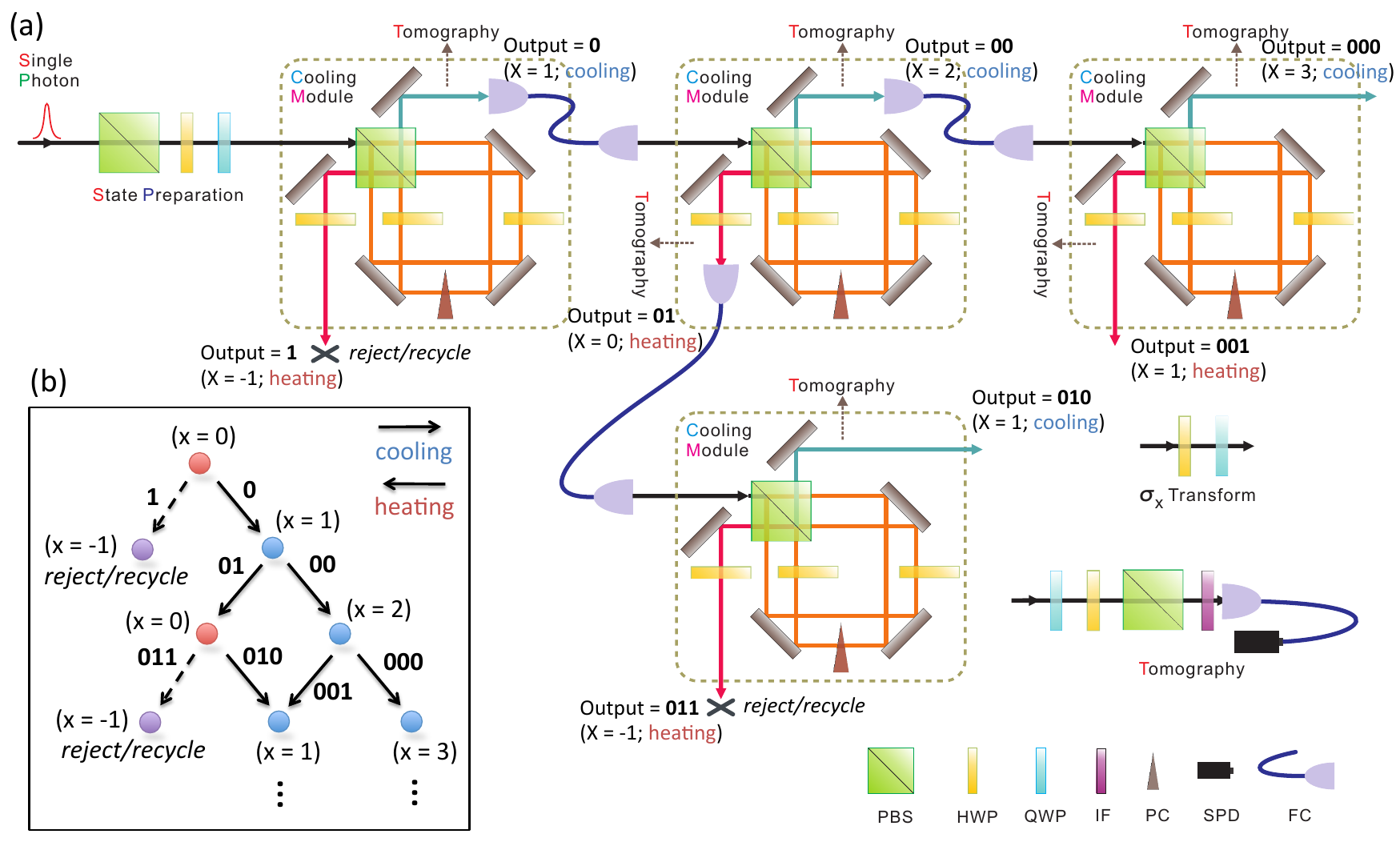}
\end{center}
\caption{(Color online). Experimental details. (a) The input photon
  state is prepared by a polarization beam splitter (PBS), a half-wave
  plate (HWP) and a quarter-wave plate (QWP). The cooling module is a
  polarization-dependent Sagnac interferometer. Split by the PBS, the
  horizontal and vertical components of a polarized photon propagate
  in opposite paths within the interferometer. Two HWPs operate on the
  corresponding polarization components. The quartz plate compensator
  (PC) is used to compensate for the relative phase in the
  interferometer. These elements simulate the system Hamiltonian. Both
  paths are then recombined on the same PBS. The photons at output
  ports $0$ and 1 represent the cases of cooling and heating,
  respectively. The photon proceeds to the next cooling module
  depending on the value of the position $x$ of the random walker. The
  final quantum state is reconstructed using quantum state tomography,
  with the measurement bases defined by a QWP, HWP and PBS. The photon
  is detected by a single photon detector (SPD) equipped with a 3 nm
  interference filter (IF). When the simulated Hamiltonian is
  $\sigma_{x}$, a HWP with an angle of $22.5^{\circ}$ and a tiltable
  QWP with an horizontal angle implement the $\sigma_x$ transformation
  at the input and output ports. (b) Sketch of the random walker
  evolution.} \label{fig:setup}
\end{figure*}

In the next step, the resulting state ${\Lambda _ \pm }\left| \psi_{in}  \right\rangle$ will either be sent to another cooling module for further cooling, or recycled/rejected (see method), depending on the measurement outcome. We formulated this procedure as a feedback-control loop as depicted in Fig.~\ref{fig:logic}a. As governed by the laws of quantum mechanics, the measurement outcomes of the ancilla qubit are intrinsically random. We model the cooling process as a 1D random walk by keeping track of the sequence of the measurement
outcomes. A net cooling effect can be achieved by implementing the feedback system that employs the following mapping: consider a random walker starts at position $x=0$. If the measurement outcome is $\left| 0 \right\rangle $, then $x$ is increased by $1$, and decreased by $1$ for $\left| 1 \right\rangle $.  When the random walker goes beyond the origin to the negative side $(x<0)$, it is either discarded (``evaporative" strategy) or recycled ( ``recycling" strategy) where the system is reset to the the initial state $\left| \psi_{in}  \right\rangle $ (see~\cite{SI} for more details). In implementing the evaporative strategy, we assumed an initial ensemble of identical states (red particles in Fig.~\ref{fig:logic}a). When the random walker position is set to $x=-1$, then the system is rejected, and we restart the experiment with another member in the ensemble. Just like ordinary evaporative cooling, this results in a decrease of overall energy but at a cost of loss of particles (blue particles in Fig.~\ref{fig:logic}a). For the recycling strategy, it is applicable to a single system. We assume the same initial state can be prepared efficiently. When the random walker position is set to $x=-1$, we simply reset the quantum state of the system to the same initial state. For $x \ne 1$, both strategies work in exactly the same way.

We note that the recycling strategy resembles the way of releasing entropy through thermalization in HBAC. The main difference is that we correlate high-entropy states with the position $(x<0)$ of the random walker instead of including a pre-assigned qubit as heat sink in HBAC. The performance and trade-off of the two strategies will be compared and explained in the experimental section. In both cases, the random walker undergoes a diffusion toward the $x>0$ direction (see Fig.~\ref{fig:setup}b). A detailed scaling analysis with Monte Carlo simulation is provided in~\cite{SI}.

In fact, the cooling module is effectively a realization of the thought experiment of Maxwell's
demon~\cite{toyabe_experimental_2010}: an intelligent being who is
able to separate hot and cool particles, which decreases the system's
entropy without any work being done on it. In our setup, ancilla
qubits are constantly refreshed, and they therefore serve as an
entropy sink. Each measurement outcome of the cooling module reveals
partial information about the eigenstates of $H_s$. If we repeatedly
apply the cooling module to the same system, the process is
asymptotically close to a quantum non-demolition
measurement~\cite{lupascu_quantum_2007} in the eigenstate basis of
$H_s$ (see~\cite{SI} for details). In other words, entropy (in the energy basis) is removed
from the system by measurements, instead of employing an external heat bath as in the heat-bath algorithmic cooling method~\cite{Boykin2002, baugh_experimental_2005}.


Below, we report a proof-of-principle demonstration of the quantum cooling method with an
all-optical setup (see e.g.~\cite{Aspuru2012} for a recent review of photonic quantum simulation). We built a network of optical elements to connect four cooling modules (see Fig.~\ref{fig:setup}a) to implement a multiple-step quantum cooling. This setup is specifically designed for cooling the polarization degrees of freedom of a single photon (see~\cite{SI} for requirements for a scalable implementation). The simulated quantum system has only two levels, i.e., a qubit. To illustrate the
flexibility of the setup, two different Hamiltonians~$H_s$ were
simulated, namely Pauli's matrices $\sigma_{z} = \left( \begin{smallmatrix} 1 &0\\ 0&-1 \end{smallmatrix} \right)$ and
$\sigma_{x}= \left( \begin{smallmatrix} 0 &1\\ 1&0 \end{smallmatrix} \right)$. We emphasize that although the Hamiltonians to be simulated are relatively simple, the ground states are not assumed to be known.  The other specifications of the simulation are as follows: the logical Hilbert space of the system was encoded in the photonic
polarization $\left\{ {\left| H \right\rangle ,\left| V \right\rangle
  } \right\}$ from a single photon source, where ${\left| H
  \right\rangle }$ and ${\left| V \right\rangle }$ respectively refer to the
horizontally and vertically polarized state of a photon. The
which-path degree of freedom $\left\{ {\left| 0 \right\rangle ,\left|
      1 \right\rangle } \right\}$ was employed as the ancilla qubit in the quantum circuit (see  Eq.~\ref{key}). The evolution operator $U\left( t \right) = \exp \left( { - iH_s t} \right)$ was simulated
for a fixed period of $t = \pi/2$. An energy offset, defined as
\begin{equation}\label{theta}
\theta \equiv \gamma + \pi/2 \quad,
\end{equation}
where $\gamma$ is defined in Fig.~\ref{fig:logic}b, was used
as an adjustable parameter for characterizing the cooling efficiency. The goal of the experiment is to minimize the mean
energy $\left\langle {\sigma _z } \right\rangle$ (or $\left\langle
  {\sigma _x } \right\rangle$) for any given initial state.

Each of the cooling modules were realized by a polarization-dependent Sagnac
interferometer (circled by dotted lines in Fig.~\ref{fig:setup}a); similar, but not identical, structures~\cite{Gao2010,Zhou2011} have been used for tasks related to quantum information processing. The quantum state $\left| \psi_{in}  \right\rangle$ of the incident photon was prepared by a series of optical elements shown in Fig.~\ref{fig:setup}a. The optical components inside the cooling module result in quantum logical
operations~\cite{Lanyon2008} which are equivalent to that described in Eq.~\ref{key} (see~\cite{SI} for details). The photon leaving from one of the two output ports of the cooling module is in a superposition state (see Eq.~\ref{key}) containing both outcomes of heating $\left| 1 \right\rangle $ (red arrow) and cooling $\left| 0 \right\rangle $ (blue
arrow). Subsequent detection of the two paths corresponds to a projective measurement in the ancilla qubit space, and will result in one of the two output states~${\Lambda _ \pm }\left| \psi_{in}  \right\rangle$.

\begin{figure}[t]
\begin{center}
\includegraphics [width= 0.85 \columnwidth]{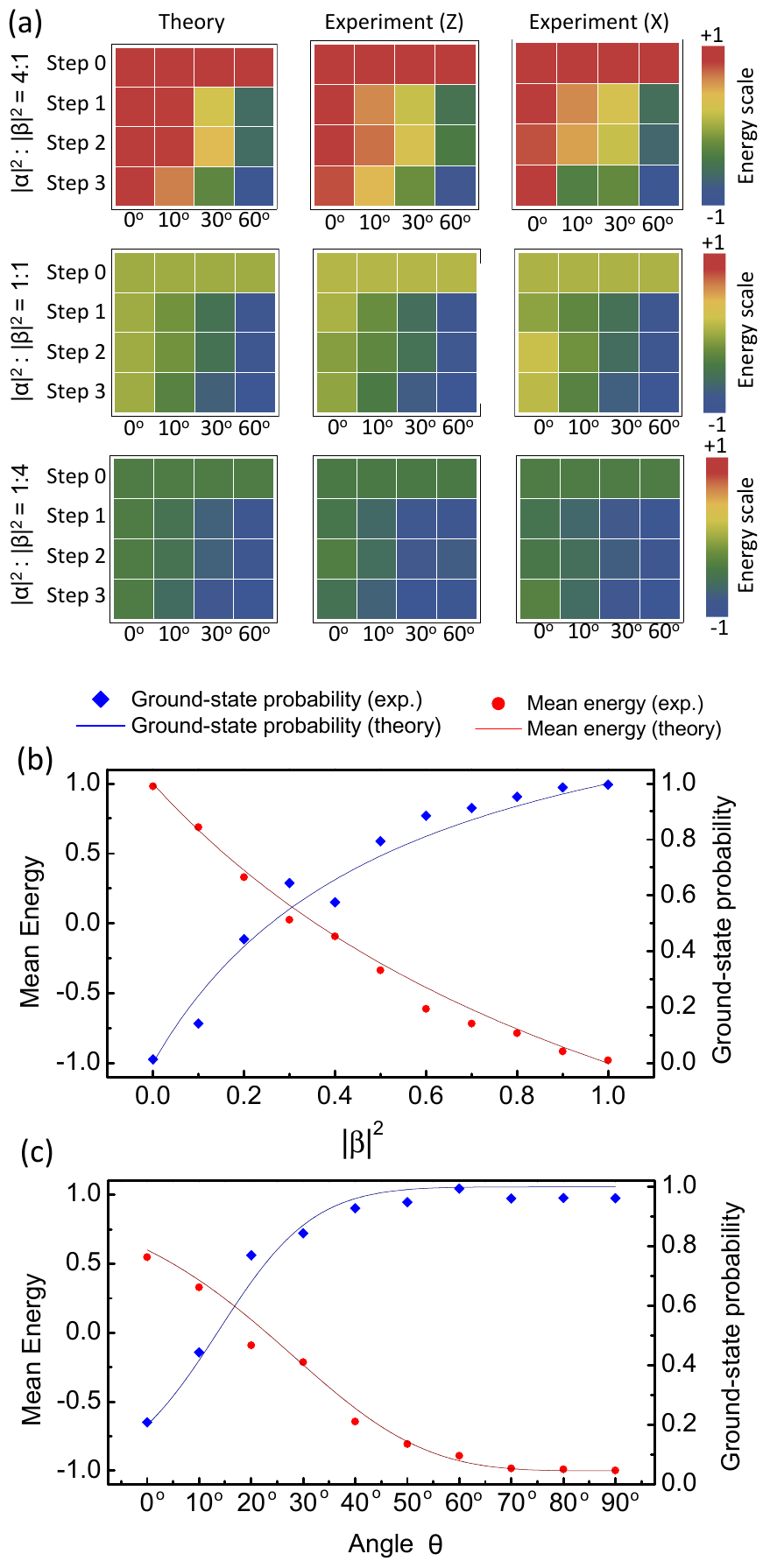}
\end{center}
\caption{(Color online). Experimental results. (a) Mean energies for 3 cooling steps of the evaporative (non-recycling) strategy for different $\theta$. Different maps correspond to theory and experiment for Hamiltonians $\sigma_{z}$ ($Z$) and $\sigma_{x}$ ($X$), and different initial states (step $0$). (b)  Theoretical and experimental mean energy and ground state probability after the 3rd cooling step as a function of the initial ground-state overlap $|\beta|^{2}$. The evaporative strategy simulated the Hamiltonian $\sigma_z$ with energy bias angle $\theta=10^{\circ}$. (c) Theoretical and experimental mean energy and ground state probability after the 3rd cooling step as a function of $\theta$. The initial input state was $(2/\sqrt{5})|e\rangle+(1/\sqrt{5})|g\rangle$ ($|\alpha|^{2}:|\beta|^{2}=4:1$).
}
\label{fig:fullcycle}
\end{figure}

For the first cooling module, a single step of cooling is achieved if the measurement outcome is $\left| 0 \right\rangle $. To further extract energy from the system, multiple steps of cooling events are needed. To proceed, four identical cooling modules are connected by optical fibers. The network of
cooling modules (as shown in Fig.~\ref{fig:setup}a) allows us to
implement both the evaporative and recycling strategies. The implementation was guided by following the random-walker diagram shown in Fig.~\ref{fig:setup}b. Specifically, for any event where the photon emerges from the cooling
(heating) exit of a cooling module, the random walk position $x$ is
increased (decreased) by $1$.  Across the boundary $x=-1$, for a range of parameters, the walkers have energy $\left\langle {H_s } \right\rangle$ higher than that of the initial state. As in evaporative cooling, separating these hot walkers from the system necessarily lowers the average energy of the system.


\begin{figure}[t]
\begin{center}
\includegraphics [width= 1 \columnwidth]{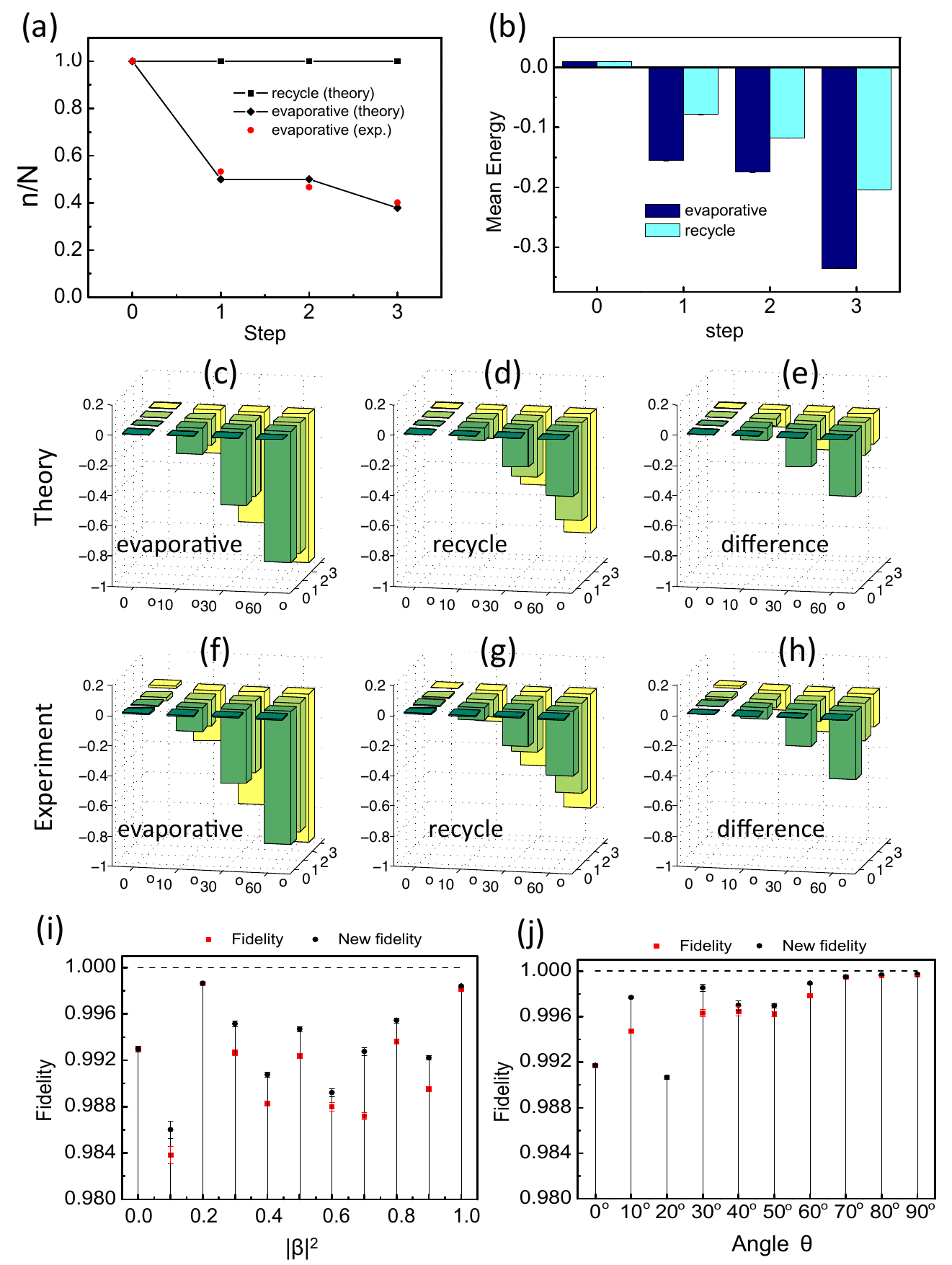}
\end{center}
\caption{(Color online). Experimental results. In (a-h) we compare the tradeoff between the evaporative and recycling cooling strategies for Hamiltonian $\sigma_{z}$, initial state $(|e\rangle+|g\rangle) /\sqrt{2}$, and bias angle $\theta=10^{\circ}$.  (a) Theory and experiment for the number $n$ of copies obtained out of $N$ initial copies at consecutive steps. The total probability of the recycling algorithm is set to $1$, correcting for experimental loses.  (b) Experimental results for the corresponding mean energies. (c-e) Theoretical results for the mean energies for the evaporative strategy, the recycling strategy, and their difference. Different bars correspond to different energy bias angles and cooling steps. (f-h) Corresponding experimental predictions. (i) and (j) are the corresponding experimental and revised fidelities. Error bars correspond to the statistical distribution of experimental measurements.}
\label{fig:fullcycle2}
\end{figure}

The experimental results with multiple steps of the cooling simulation
are summarized in Fig.~\ref{fig:fullcycle}.  The color-coded energy
maps of Fig.~\ref{fig:fullcycle}a correspond to the evaporative
strategy. To characterize the efficiency of the cooling method, we considered different initial input states~$\alpha|e\rangle+\beta|g\rangle$ for Hamiltonians $H_s = \sigma _z$ and $ \sigma _x $, and different proportions $|\alpha|^{2}:|\beta|^{2}$ between the populations of the ground $\left| g \right\rangle$ and
excited states~$\left| e \right\rangle$. We remark that (i) the cooling method does not require the knowledge of the ground states, and (ii) the theoretical values of the mean energy are the same for both Hamiltonians. The first row in each
color coded energy map (step $0$) represents the mean energy of the
initial state. The other rows in an energy map correspond to the mean
energy after each step of cooling. The columns of the energy maps
correspond to different choices of the energy bias $\theta \equiv
\gamma+ \pi/2$ (cf. Eq.~\ref{theta}) in the cooling module. For an angle $\theta = 0^\circ$, the mean energy remains constant.  For other angle settings
($\theta \ne 0^o$), the mean energy decreases with increasing cooling
steps. Note that in the evaporative strategy, after the first step, the output states corresponding
to heating events ($x=-1$) are discarded. This is reflected in the
fact that for all angles $\theta \ne 0^o$ the mean energy decreases
(output port $0$ of the first cooling module). However, at the second
step, the random walkers jump to either $x=0$ and $x=2$ and no
walker is rejected (output ports $00$ and $01$, see also
Fig.~\ref{fig:setup}b). Consequently, the mean energies after the
first and second steps are the same (see also
Fig.~\ref{fig:fullcycle2}a).

We systematically probed the mean energy $\left\langle {H_s } \right\rangle$ and the ground-state probability $P_g  = \left\langle g \right|\rho _{\exp } \left| g \right\rangle$ of the output states corresponding to the cooling events after three steps, Figs.~\ref{fig:fullcycle}b and~\ref{fig:fullcycle}c. The mean energy decreases with increasing ground-state probability $\left| \beta \right|^2$ and energy bias $\theta$. The output state
approaches the ground state when $\theta$ is close to $90^o$, in good
agreement with the theoretical prediction. In our experiment, the
statistics of each count are considered to follow a Poisson
distribution and the error bars are estimated from the standard
deviations of the values calculated by the Monte Carlo method.  We compared the trade-off between the strategies with and without
recycling of ``hot'' copies. \mbox{Figures~\ref{fig:fullcycle2}(a-h)} show the
theoretical and experimental results with initial input state
$(|e\rangle+|g\rangle)/\sqrt{2}$, Hamiltonian $\sigma_{z}$, and
energy bias angle $\theta = 10^\circ$. It is clear that the mean energy of the evaporative strategy decreases faster than that of the recycling strategy at the cost of a lower total yield. Finally, the
experimental fidelities are always larger than 0.983 and the
fidelities after post-processing (see methods summary) are even
better (compare black vs red data points in Fig.~\ref{fig:fullcycle2}i and \ref{fig:fullcycle2}j).

To conclude, our simulation results suggest that cooling for quantum simulation can be experimentally achieved with high fidelity. Due to the modest experimental resource requirement, even the fully scalable version (see~\cite{SI}) of this method can potentially be implemented with many of the currently available quantum computing architectures, such as trapped ions, superconducting devices and NMR systems. Furthermore, as the solution to classical~\cite{Farhi2001} or quantum~\cite{Feynman1985} computational problems can be encoded in the ground state of certain Hamiltonian, our cooling method can potentially have applications beyond quantum simulation. 


\section{Methods}

Here we justify the claim in Eq.~\ref{key} that for any quantum state $\left| {{\psi _{in}}} \right\rangle $ that is not an eigenstate of $H_s$, the state ${\Lambda _ - }\left| \psi_{in}  \right\rangle$ (${\Lambda _ + }\left| \psi_{in}  \right\rangle$), after normalization, has energy $E = \left\langle H_s \right\rangle $ lower (higher) than that of $\left| {{\psi _{in}}} \right\rangle $. To proceed, we performed a eigenvector expansion, which gives $\left| {{\psi _{in}}} \right\rangle  = \sum\nolimits_k {{c_k}} \left| {{e_k}} \right\rangle $, where $\left| {{e_k}} \right\rangle$'s are the eigenvectors of $H_s$, i.e., $H_s\left| {{e_k}} \right\rangle  = {E_k}\left| {{e_k}} \right\rangle$, and ${c_k} = \left\langle {{e_k}\left| {{\psi _{in}}} \right\rangle } \right.$. Consider ${\Lambda _ \pm }\left| {{\psi _{in}}} \right\rangle  = \sum\nolimits_k {{c_k}\left( {1 \pm i{e^{ - i{\phi _k}}}} \right)\left| {{e_k}} \right\rangle }$, where ${\phi _k} \equiv {E_k}t - \gamma $. Note that the square norm of the amplitudes are ${\left| {{c_k}} \right|^2}\left( {1 \pm \sin {\phi _k}} \right)$. It means that apart from an overall constant, each of the weight ${\left| {{c_k}} \right|^2}$ of the eigenvectors are scaled by a factor $\left( {1 \pm \sin {\phi _k}} \right)$. In Fig.~\ref{fig:logic}c, within the range $\left| {{\phi _k}} \right| < \pi /2$, the function ${1 - \sin {\phi _k}}$ is monotonically decreasing. In other words, the lower the eigen-energy is, the higher the weight gains, which results in a lower average energy, i.e., cooling. A similar argument applies for the case of heating.

Here we explain the difference between the ``evaporative" strategy and the ``recycling" strategy. In implementing the evaporative strategy, we assumed an initial ensemble of identical states (red particles in Fig.~\ref{fig:logic}a). When the random walker position is set to $x=-1$, then the system is rejected, and we restart the experiment with another member in the ensemble. Just like ordinary evaporative cooling, this results in a decrease of overall energy but at a cost of loss of particles (blue particles in Fig.~\ref{fig:logic}a). For the recycling strategy, it is applicable to a single system. We assume the same initial state can be prepared efficiently. When the random walker position is set to $x=-1$, we simply reset the quantum state of the system to the same initial state. For $x \ne 1$, both strategies work in exactly the same way.

The revised fidelities in Fig.~\ref{fig:fullcycle}d and \ref{fig:fullcycle}e are obtained by mapping each experimental
density matrix (obtained using maximum likelihood) to its eigenvector
with highest eigenvalue. This corrects isotropic (depolarizing) noise.

\section*{Author Contributions}
M.-H.Y. and A.A.-G. are responsible for the main theoretical idea of the cooling method. S.B. performed detailed analysis on the scaling behavior of cooling method. M.-H.Y. drafted the preliminary experimental proposal which was put forwarded by Z.-W.Z. and C.-F.L. The detailed experimental procedures were designed and carried out by J.-S.X., assisted by X.-Y.X. The experiment was supervised by C.-F.L. and G.-C.G. All authors contributed to the writing of the paper and discussed the experimental procedures and results.

\section{Acknowledgments}
We thank A. Eisfeld, I. Kassal, J. Taylor, and J. Whitfield for insightful discussions, and R. Babbush, S. Mostame, and D. Tempel for useful comments and suggestions on the manuscript. We are grateful to the following funding sources: National Science Foundation award no. CHE-1037992, Croucher Foundation (M.H.Y); DARPA under the Young Faculty Award N66001-09-1-2101-DOD35CAP, the Camille and Henry Dreyfus Foundation, and the Sloan Foundation; DARPA award no. N66001-09-1-2101 (S.B.); the National Basic Research Program of China (Grants No. 2011CB9212000), National Natural Science Foundation of China (Grant Nos. 11004185, 60921091, 10874162, 10874170), and the Fundamental Research Funds for the Central Universities (Grant No. WK 2030020019). 

\section{Additional information}
The authors declare no competing financial interests.



\input{SI}

\end{document}

%% file: SI.tex
\clearpage

\onecolumngrid


\pagestyle{empty}

\section{Supplementary Information}

\vspace{5 mm}
\centerline {\bf Demon-like Algorithmic Quantum Cooling and its Realization with Quantum Optics}
\vspace{5 mm}
\centerline{Jin-Shi Xu, Man-Hong Yung, Xiao-Ye Xu, Sergio Boixo,} 
\centerline{Zheng-Wei Zhou, Chuan-Feng Li, Al\'{a}n Aspuru-Guzik, Guang-Can Guo}
\vspace{5 mm}



\tableofcontents

\newpage

\section{1: Basic principle of the simulated cooling algorithm}
Here we explain in detail the basic idea of the cooling algorithm implemented in this experiment. As shown in Fig.~\ref{fig:logic}b, the core of the quantum circuit (cooling module) consists of four components: (a) two Hadamard gates
\begin{equation}
\mathsf{H} \equiv {\textstyle{1 \over {\sqrt 2 }}}\left( {\left| 0 \right\rangle  + \left| 1 \right\rangle } \right)\left\langle 0 \right| + {\textstyle{1 \over {\sqrt 2 }}}\left( {\left| 0 \right\rangle  - \left| 1 \right\rangle } \right)\left\langle 1 \right| \; ,
\end{equation}
applied to the ancilla qubit in the beginning and at the end of the quantum circuit. (b) a local phase gate,
\begin{equation}
\mathsf{R_z} \left( \gamma  \right) \equiv \left| 0 \right\rangle \left\langle 0 \right| - ie^{i\gamma } \left| 1 \right\rangle \left\langle 1 \right| \;,
\end{equation}
where the parameter $\gamma$ plays a role in determining the overall efficiency of the cooling performance of the algorithm. (c) a two-qubit controlled unitary operation,
\begin{equation}
\openone \otimes \left| 0 \right\rangle \left\langle 0 \right| + U \otimes \left| 1 \right\rangle \left\langle 1 \right| \;,
\end{equation}
where $\openone$ is the identity operator, and
\begin{equation}
U\left( t \right) = e^{ - i H_s t}
\end{equation}
is the time-evolution operator for the system. Here $H_s$ is the
Hamiltonian of the system, and the energy is defined as the quantum
expectation value of $H_s$. The parameter $t$ also determines the
overall efficiency of the algorithm.

For any given initial state (it works equally well for mixed states), $\left| {\psi _{in} } \right\rangle$, the algorithm starts with the product state of the system state $\left| {\psi _{in} } \right\rangle$ and the ancilla state $ \left| 0 \right\rangle$,
\begin{equation}
\left| {\psi _{in} } \right\rangle  \otimes \left| 0 \right\rangle \;.
\end{equation}
The quantum circuit then produces the following output state:
\begin{equation}\label{core_step}
\frac{1}{2}\left( {\openone - ie^{i\gamma } U} \right)\left| {\psi _{in} } \right\rangle \left| 0 \right\rangle  + \frac{1}{2}\left( {\openone + ie^{i\gamma } U} \right)\left| {\psi _{in} } \right\rangle \left| 1 \right\rangle \;.
\end{equation}
A projective measurement on the ancilla qubit in the computational basis $\left\{ {\left| 0 \right\rangle ,\left| 1 \right\rangle } \right\}$  yields one of the two (unnormalized) states
\begin{equation}
\left( {\openone \pm ie^{i\gamma } U} \right)\left| {\psi _{in} } \right\rangle,
\end{equation}
which have mean energies either higher (for outcome $\left| 1 \right\rangle$) or lower (for outcome $\left| 0 \right\rangle$) than that of the initial state $\left| {\psi _{in} } \right\rangle $.

To justify this statement, let us expand the input state in the eigenvector basis $\left\{ {\left| {e_k } \right\rangle } \right\}$ of the Hamiltonian $H_s$,
\begin{equation}
\left| {\psi _{in} } \right\rangle  = \sum\limits_k {c_k } \left| {e_k } \right\rangle \;.
\end{equation}
Note that the vector norms have the following form:
\begin{equation}
\left| {\left( {1 \pm ie^{i\gamma } U} \right)\left| {e_k } \right\rangle } \right|^2  =  2 \left( {1 \pm \sin \phi _k } \right) \;,
\end{equation}
where $\phi _k  \equiv E_k t - \gamma$ depends on the eigenenergy $E_k$ of $H_s$. In other words, each of the eigenvector weights $|c_k|^2$ is now multiplied by a factor
\begin{equation}
1 \pm \sin \phi _k
\end{equation}
depending on the measurement result. To simplify our discussion, we will assume that one can always adjust the two parameters, $\gamma$ and $t$, such that
\begin{equation}
- {\textstyle{\pi  \over 2}} \le \phi _k  < {\textstyle{\pi  \over 2}}
\end{equation}
for all $k$'s. Then, the factors $(1 - \sin \phi _k)$ are in descending order of the eigenenergies (see Fig. \ref{fig:logic}c), and the opposite is true for the factors $(1 + \sin \phi _k)$ .

Therefore, apart from an overall normalization constant, the action of the operator $({\openone \pm ie^{i\gamma } U})$ is to scale each of the probability weight $\left| {c_k } \right|^2$ by a factor of $(1 \pm \sin \phi _k)$; the probability weights  scale to larger values,  i.e.,
\begin{equation}
\left( {1 - \sin \phi _k } \right)/\left( {1 - \sin \phi _j } \right) > 1
\end{equation}
for the eigenenergy $E_k < E_j$ in the cooling case (i.e., for outcome $\left| 0 \right\rangle$), and vice versa for the heating case (i.e., for outcome $\left| 1 \right\rangle$).

\subsection{Comments about the measurement}

The one-bit measurement result of the cooling module correlates with
the energy of the state of the simulated system. After the measurement, the state of the system changes. If the state of the system is diagonal in the energy basis, this is strictly a Bayesian update. In general, this is nothing but a (partial) ``collapse'' of the wave function. If the measurement result is 0, the relative population of the low energy eigenstates of the system increases, Fig.~\ref{fig:logic}(c). That is, if we post-select copies for which
the results of a sequence of measurements have a sufficiently higher
proportion of zeros than ones, then the average energy decreases, as
in evaporative cooling.  

When each measurement corresponds to a
short time evolution this technique is related to imaginary time
evolution, as in diffusion Monte Carlo \cite{aspuru2003}. If we do not post-select and the system is finite,
the relative frequency of zeros and ones will peak on well defined
values for different copies of the system. These well defined values are in a one-to-one
correspondence with the possible energies of the Hamiltonian.
The measurement of the energy can be viewed as a simulation of von Neumann's description of the quantum measurement process: the evolution of the system with Hamiltonian $H$ couples with the position of a ``measuring pointer.'' The pointer, in our case, is a binary ancilla qubit. 

\subsection{Comments about ``cooling"}
With our method, we do not result directly with a thermal state with Boltzmann distribution. Therefore, temperature is not well-defined. However, the entropy associated with the diagonal elements in the energy basis can be properly defined. Our method allows us to reduce the average energy and also the entropy. Of course, it is also possible to combine our method with other existing methods to project the exact thermal states, which will also allow us to define temperature. 

For example, if we start with the identity matrix $I$ as our initial state, we can consider it as an infinite-temperature state. When we simulate a small time evolution and zero $\gamma$, then the scaling factor (e.g. see Fig.~\ref{fig:logic}) becomes approximately the same as the Boltzmann factor
\begin{equation}
1 - \sin {\phi _k} \approx {e^{ - {E_k}t}} \quad.
\end{equation}
To further improve the fidelity, one may apply the projection method described in Ref.~\cite{poulin_preparing_2009, bilgin_preparing_2010}. We can repeat the same procedure to reach states with a lower temperature.

\subsection{Requirement for a fully scalable implementation}
\begin{figure}[t]
  \centering
  \includegraphics[width=0.9 \columnwidth]{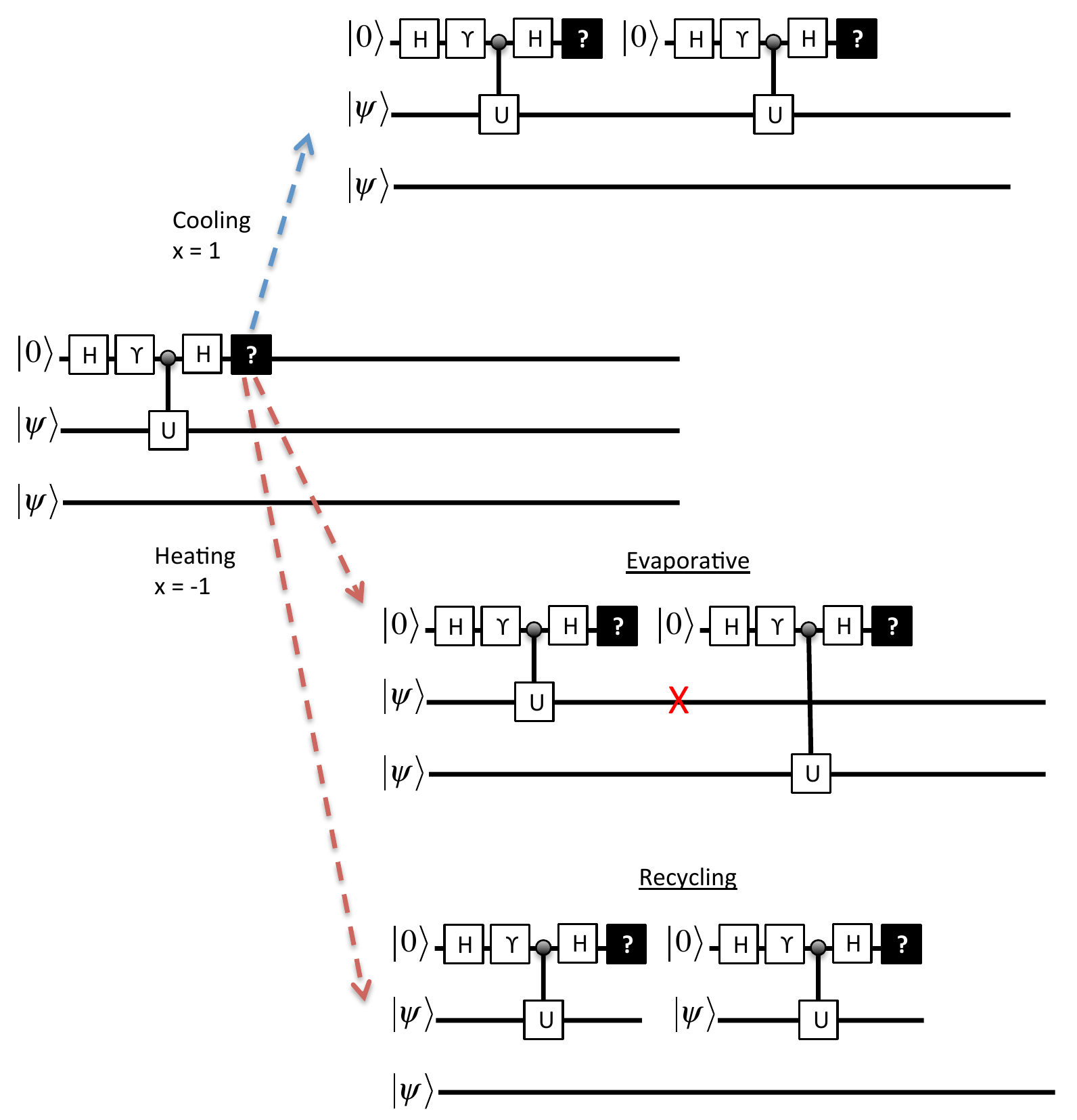}
  \caption{An example of the scalable implementation of the quantum cooling algorithm. When the first step of cooling is implemented, a projective measurement (labeled by a question mark ``?") is performed on the first ancilla qubit. The measurement result then determines if the second ancilla would interact with the first system or the second system. If cooling is resulted, then $x=1$, and the same quantum circuit is applied to the same quantum system. It is same for both evaporative strategy and recycling strategy. However, if the measurement outcome is heating, then for the evaporative strategy, the system is rejected, and a new member is employed. For the recycling strategy, the same system is reset and sent to the same quantum circuit.}
\label{fig:scalable}
\end{figure}
In our experiment, we demonstrated a proof-of-principle experiment for implementing our cooling algorithm. However, the experimental setup is specifically designed for cooling the polarization degrees of freedom. Here we discuss the requirements for a fully scalable implementation. First of all, we need a supply of fresh ancilla qubits prepared in the $\left| 0 \right\rangle$ state. It is not strictly required to have perfect $\left| 0 \right\rangle$ ancilla, but an error would decrease the cooling efficiency. Alternatively, if we are able to reset the ancilla qubit quickly, we can also employ only one ancilla qubit through out. Secondly, we will need to be able to maintain the coherence of the system qubits. Therefore, we expect error corrections will still be required, unless we can finish the cooling within the coherence time of the system. Thirdly, we will also need to have fast projective measurement on the ancilla qubits, which will allow us to be able to implement our feedback control system. An example is shown in Fig.~\ref{fig:scalable}. These requirements are not much different from those required for implementing a typical quantum algorithm. The main feature of this cooling algorithm is the modest requirement of using ancilla qubits, which could reduce the resources for expanding it into a fault-tolerant structure.

\section{2. Scaling analysis of the simulated cooling algorithm}
\subsection{Simplified model}
We study the complexity of preparing the ground state for easy of
comparison with other methods. We focus on Hamiltonians with only two
different energies. The generalization of the same problem to systems
with more energies is straightforward. The degeneracy of the energy
levels turns out not to be of any consequence for the resulting cost,
only the projection probability on the lowest energy subspace is
important. With that in mind, we will write the equations for a two
level system to simplify the notation.

We denote the two energy eigenstates by $\ket{e_0}$ and $\ket{e_1}$
with energies $E_0$ and $E_1$, and define 
\begin{equation}
\left| {\psi _{in} } \right\rangle  = \sqrt p \left| {e_0 } \right\rangle  + \sqrt {1 - p} \left| {e_1 } \right\rangle \quad. 
\end{equation}
Let us ignore for the moment the
possibility of filtering high energy states after too many ``heating''
measurements. The insight gained with this assumption can be applied
to the more sophisticated algorithms in the text. We can then also
assume, without loss of generality, that all the measurements are done
together at the end of the computation. Therefore, if there are a
total of $k$ steps, we get the state (prior to the final measurement)

\begin{equation}
  \sum_{j=0}^k \sqrt{ \binom k j } \Big( \sqrt{p (1/2+a)^{k-j} (1/2-a)^j} \ket {e_0} \ket {\mathcal X_j} \nonumber +   \sqrt{(1-p) (1/2+b)^{k-j} (1/2-b)^j} \ket{e_1} \ket{\mathcal Y_j}\Big) \;,
\end{equation}
where $\ket {\mathcal X_j}$ and $\ket {\mathcal Y_j}$ are properly
normalized states, 
\begin{equation}
a = \frac 1 2 \sin (E_0 t - \gamma) \quad , \quad b = \frac 1 2 \sin (E_1 t - \gamma) \quad, 
\end{equation}
and $j$ denotes the numbers of $0$'s in
both $\ket {\mathcal X_j}$ and $\ket {\mathcal Y_j}$. The parameter
$\gamma$ is a controllable energy bias, see Fig.~\ref{fig:logic} in the
main text.

The probability to get $j$ zeros on the final measurements of the ancillae is

\begin{equation}
p  \binom k j  (1/2+a)^{k-j} (1/2-a)^j \nonumber + (1-p) \binom k j  (1/2+b)^{k-j} (1/2-b)^j \; .
\end{equation}
This is a mixture (with mixing probability $p$) of two binomial
distributions. The left side binomial (corresponding to the ground
state) concentrates on the interval
\begin{align}
  k \left( \left(\frac 1 2 -a\right) \pm \frac 1 {\sqrt k} \sqrt{\frac 1 4 - a^2} \right)
\end{align}
and the right side binomial concentrates on the interval
\begin{align}
   k \left( \left(\frac 1 2 -b\right) \pm \frac 1 {\sqrt k} \sqrt{\frac 1 4 - b^2} \right)\;.
\end{align}

The first binomial corresponds to the probability of finding the
ground state in the post-condition state after getting $j$
zeros. Because we want to prepare the ground state, the binomials
should have little intersection. This happens when

\begin{equation}
  k \left( \left(\frac 1 2 -b\right) + \frac 1 {\sqrt k} \sqrt{\frac 1 4 - b^2} \right) \nonumber  \ll  k \left( \left(\frac 1 2 -a\right) - \frac 1 {\sqrt k} \sqrt{\frac 1 4 - a^2} \right) \; .
\end{equation}
This is equivalent to
\begin{align}
  \sqrt k (b-a) \gg \frac 1 2 - (a^2 + b^2)\;.
\end{align}
The most relevant case is when the energies and the bias are small. We can do an expansion in terms of the small parameter $\Delta t$, where $\Delta$ is the gap and $t$ is the evolution time. The parameters $a$ and $b$ are
then small, $\cO(\Delta t)$, using big-O notation. We get
\begin{align}
  \frac 1 2 - (a^2 + b^2) =     \frac 1 2 + \cO((\Delta t)^2) \\
  b - a = \frac {\Delta t} 2 + \cO((\Delta t)^2)\;.
\end{align}
This gives the expected result
\begin{align}\label{eq:scaling}
  \sqrt k \gg \frac 1 {\Delta t}\;.
\end{align}
That is, the number of measurements necessary to prepare the ground states has an scaling
\begin{align}
  k \in \cO((\Delta t)^{-2})\;.
\end{align}

If the binomials are separated, when we measure we get a number of zeros 
\begin{equation}
j \sim k(1/2 - b) \quad,
\end{equation}
which happens with probability $1-p$, and then we have collapsed on $\ket {e_1}$, or we get a number of 
\begin{equation}
j \sim k(1/2 -a) \quad,
\end{equation}
which happens with probability $p$, and then we have collapsed on $\ket {e_0}$. The number of measurements necessary to prepare the ground state scales like
\begin{align}
  \frac 1 {p (\Delta t)^2}\;.
\end{align}

\subsection{Monte Carlo simulations}

\begin{figure}[t]
  \centering
  \includegraphics[width=0.4 \columnwidth]{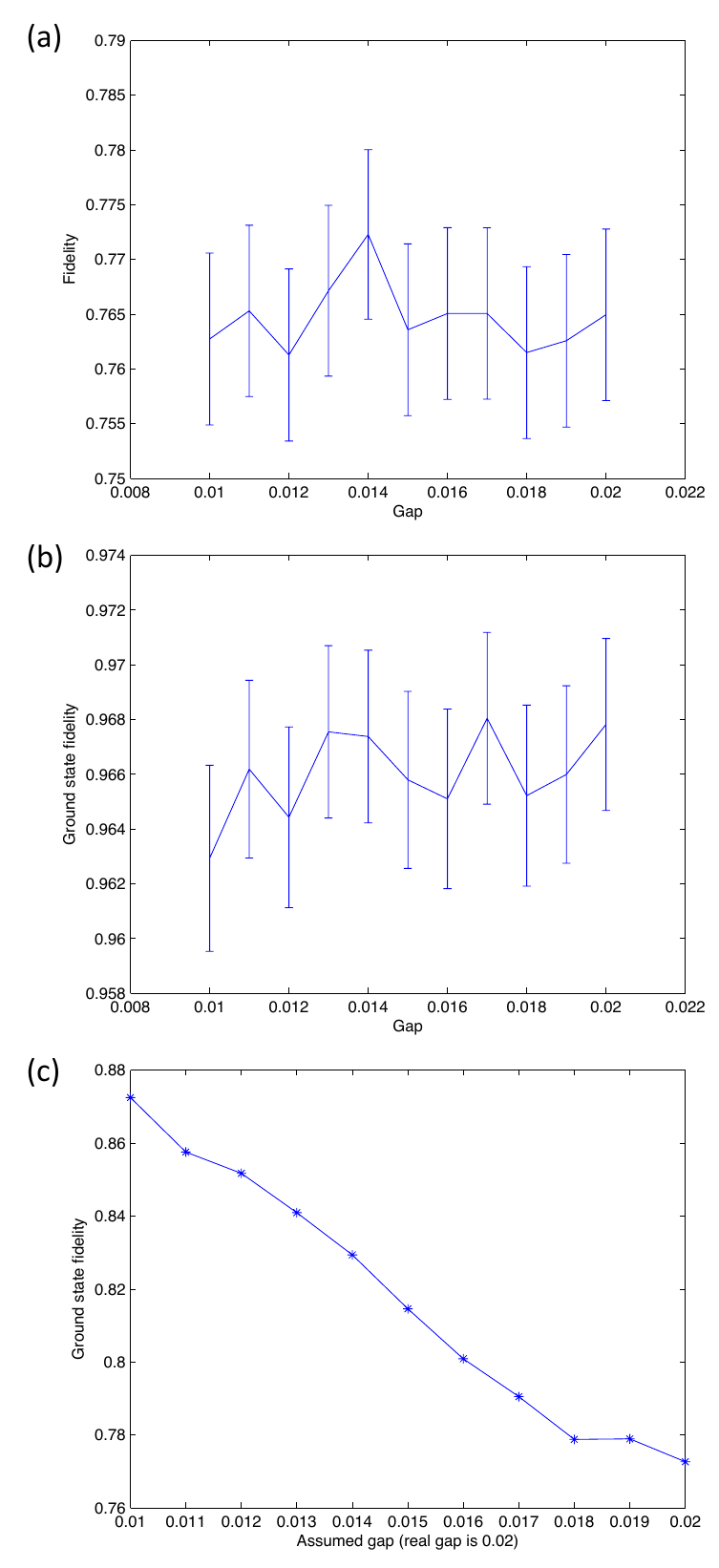}
  \caption{(a) Final fidelity for the ground state for the values explained in the text. (b) Final fidelity for the ground states for the values explained in the text when trowing away states that reach $c_{\rm bound}$ measurements. (c) Fidelity changes if we choose the reflective and absorbing boundary conditions according to the equations in the text, but using assumed gap values that do not corresponding to the real gap. The gap was kept at $0.02$, but the gap used for the boundary conditions was the one indicated in the $x$-axis.}
\label{fig:temperature_reflective}
\end{figure}

We performed Monte Carlo simulations to test the scalings explained
above for better algorithms, such as those implemented in the
experiment. Notice that in the previous subsection we obtained
scalings for the case when all the measurements are performed at the
end or, equivalently, measurements are taken at each step but the
result of the measurement is not taken into account at subsequent
steps. Nevertheless, in the experiment adaptive action was taken
taking into account the results of previous measurements (absorption
and reflection boundaries, explained in main text). The intuition is
that even an optimal adaptive action does not change the fundamental
scaling. This is what we saw with Monte Carlo simulations.

The actual experiment implemented an algorithm that keeps track of the
difference $x$ between the number of measurement results equal to $0$
and those equal to $1$ in the ancilla, 
\begin{equation}
x = \#0 - \#1 \quad.  
\end{equation}
The algorithm implemented in the experiment obeys two boundary conditions:
\begin{enumerate}
\item If the number of measurement results equal to $0$ is less than
  the number of $1$'s ($x\le0$), we reinitialize with the initial
  state. This is a ``reflective'' boundary condition.
\item For some predefined constant $c_{\rm abs}$, if $x = c_{\rm abs}$
  then stop. This is an ``absorbing'' boundary condition.
\end{enumerate}
In addition, we need to define a constant $c_{\rm bound}$ such that we
stop after reaching this number of steps. This is because this simple
method can collapse to the excited state and never reach the absorbing
wall (see previous subsection).

We can analyze this algorithm following the intuition gained from the
non-adaptive case explained above. Because we are interested in the
asymptotic case with small energy differences, we set the energy bias
\begin{equation}
\gamma = 0 \quad.
\end{equation}
The position $x$ of the random walker without boundary
conditions was found to be distributed as a mixture of two (displaced)
binomials distributions, centered around
\begin{align}
  \sin(E t) * k
\end{align}
with $E$ taking the values of the ground and excited state energies,
$k$ the total number of measurements, and $t$ is the evolution
time.

For the constant $c_{\rm bound}$ we can use our prediction on the number of steps needed to cool optimally
\begin{align}\label{eq:fundamental_scaling}
  c_1 \frac 1 {(\Delta t)^2} \frac 1 p\;,
\end{align}
with a constant $c_1$ that depends on the details of the algorithm. We
want to show that this scaling holds for significant changes in
$\Delta t$ and $p$, when this parameters have small values.  Because
our process is not optimal, we actually use
\begin{align}
  c_{\rm bound} &= 3 * {\rm pred} \\ {\rm pred} &= c_1 \frac 1 {\Delta^2 p} \;,
\end{align}
where we use the constant $c_1 \sim 2.7$ from the numerics of the
optimal filtering algorithm (see below). Although in principle one can
only use estimates of the gap $\Delta$ and the initial population $p$,
we assume that this estimates are indeed good, and use the exact
values in our numerics. Furthermore, we set the ground state energy
$E_0 = -\Delta$ and the excited state energy $E_1 = 0$. These choices
do not affect the fundamental scaling.

Finally, we choose a value for the absorption boundary condition. We
want this to be a value for $x$ where states already have a good
overlap with the ground state. We simply use the predicted average
after the predicted number of measurements necessary to collapse
\begin{align}
  c_{\rm abs} = {\rm pred}*  \sin(E_0 t)\;.
\end{align}

We performed Monte-Carlo numerics for $p=0.2$ and $\Delta t$ between
$0.01$ and $0.02$, Fig.~\ref{fig:temperature_reflective}. For these
values, the number of maximum measurements ($c_{\rm bound}$) goes
between $400000$ and $100000$, and the positions of the absorbing
boundary condition goes between $1350$ and $675$. Nevertheless, the
final fidelity with the ground state is quite constant, a clear
indication that we understand the scaling correctly. The fidelity is
always close to $0.76$. For the error bars we use the expression
\begin{align}\label{eq:error_bar}
  \hat \mu \pm 1.96 \frac{\hat \sigma}{100}\;,
\end{align}
where $\hat \sigma$ is the sample standard deviation, $\hat \mu$ is
the sample average, and we divide by the square root of the number of
samples, $\sqrt{10000}$.

As a simple improvement, we can remove from the calculation of the
fidelity the states that reach the maximum number of steps, $c_{\rm
  bound}$. This corresponds to simply restarting again with an initial
state. With the values chosen, only $20\%$ of the measurement
histories arrive to $c_{\rm bound}$, so the probability of cooling
with this method is also high. Further, the probability to always
arrive at $c_{\rm bound}$ decreases exponentially in the number of
trials. The corresponding fidelities are depicted in
Fig.~\ref{fig:temperature_reflective}.

We also check that the fidelity does indeed vary substantially if we keep the gap fixed but  change the reflective and absorbing bound conditions. This is plotted in Fig.~\ref{fig:temperature_reflective}. For these numerics the gap was kept constant at $0.02$, but the reflective and absorbing boundary conditions were chosen as if the gap was different (between $0.01$ and $0.02$, as before).

We also tested that the same fundamental scaling holds for an optimal
refreshing schedule. The optimal refreshing schedule for the two level
system is this: replace the current state with the initial state
whenever the temperature of the current state is higher than the
temperature of the initial state. This can be calculated with linear
cost in the number of steps assuming that the energies of the ground
state and the excited states are known, as well as the initial
population of the ground state. In this case the boundary conditions
and the bound on the maximum number of steps are
unnecessary. Nevertheless, the distribution of the number of steps was
found to be given by Eq.~\eqref{eq:fundamental_scaling}. The value of
the constant $c_1$ was found by fitting the Monte Carlo sampling data.

\subsection{Summary}
In summary, we focus on the complexity of preparing the ground state.  The scaling is
expressed in terms of the energy gap $\Delta$ between the ground state
and the first excited state, the evolution time $t$ of the simulated
Hamiltonian for the weak energy measurement, and the population
(probability) $p$ of the ground state. The basic scaling is inversely
proportional to $(\Delta\, t)^2 p$.  If possible, the theoretical
optimal scaling occurs when $t \sim 1/\Delta$, giving single-ancilla
quantum phase estimation
. Longer evolution times for each measurement require longer coherent times for
the ancillae. It is also worth noting that the post-conditioned distribution is different for different evolution times
and energy biases, which might be advantageous in some situations. Finally, in theory, the scaling with $p$ can be improved to $\sqrt p$ using amplitude amplification
. This comes at the expense of many more quantum operations, and therefore is less practical than the approach proposed here.

\section{3. From quantum circuit to experimental implementation}
\begin{figure*}[t]
\begin{center}
\includegraphics [width= 0.9 \columnwidth]{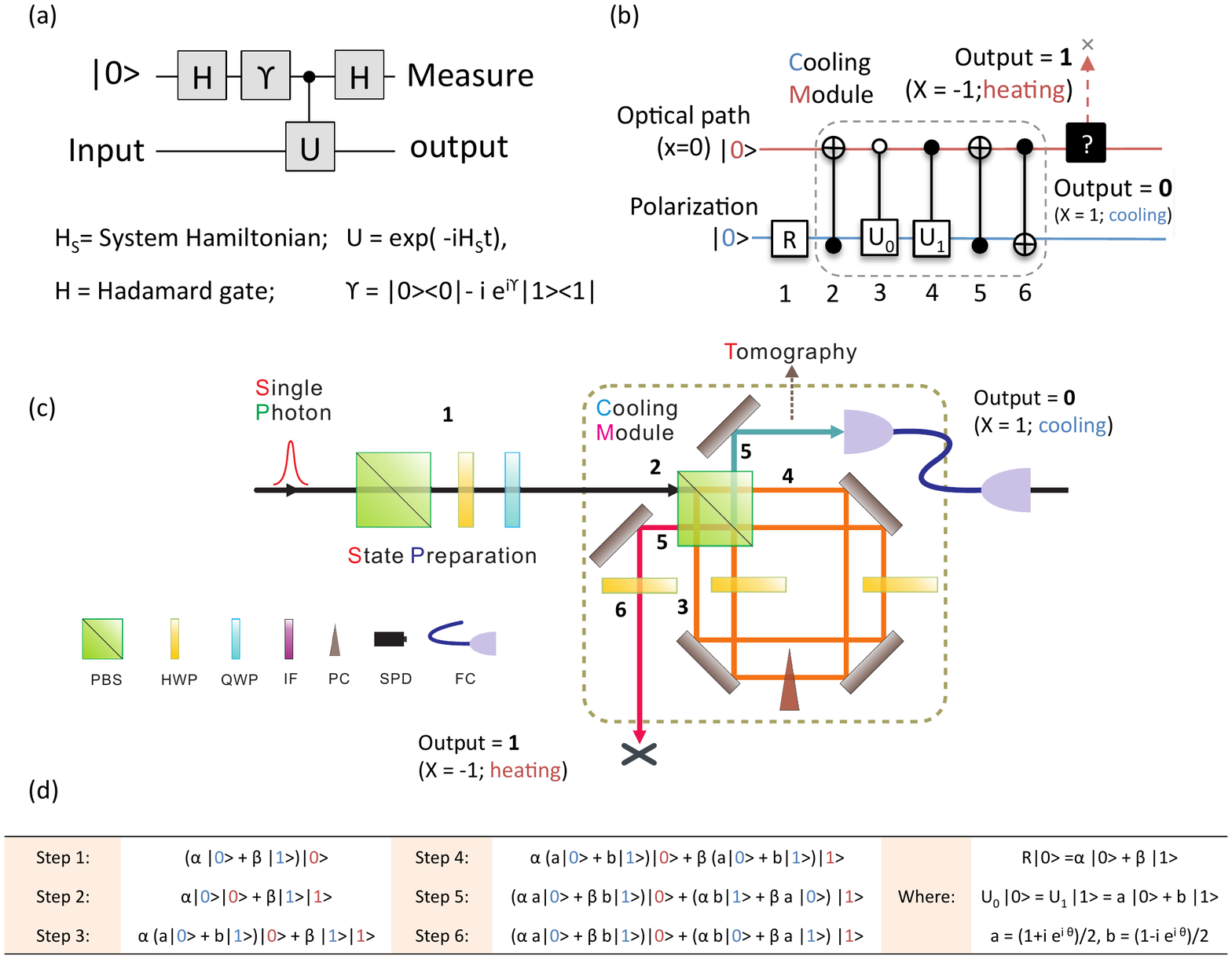}
\end{center}
\caption{(Color online). The conversion from the quantum circuit to the experimental implementation. The abbreviations refers to polarization beam splitter (PBS), half-wave plate (HWP), quarter-wave plate (QWP), interference filter (IF), quartz plate compensator (PC), single photon detector (SPD), and fiber connector (FC). See the text for the explanation of the transformations.}
\label{fig:q_circuit2exp}
\end{figure*}
In this section, we describe the experimental implementation of the quantum circuit for the cooling module in Fig.~{\ref{fig:logic}}. We will focus on a single cooling module, and the cooling with respect to the Hamiltonian $\sigma_z$. The $\sigma_x$ Hamiltonian is achieved by a rotation of the basis. Here we reproduce the relevant parts of the figures in the main text in Fig.~\ref{fig:q_circuit2exp}. The goal here is to explain how to achieve a unitary transformation that is equivalent to the one in the quantum circuit shown in Fig.~\ref{fig:q_circuit2exp}a. Our strategy is not to implement each operation separately, but to construct an effective transformation that produces the same overall result as the quantum circuit. 

\subsection{Specification of certain simulation details}
\begin{enumerate}

\item In our experiment, for the Hamiltonian $\sigma_{z}$, the mapping of the excited $|e\rangle$ and ground state $|g\rangle$ is $|e\rangle=|H\rangle$ and $|g\rangle=|V\rangle$, where $|H\rangle$ and $|V\rangle$ represent the horizontal and vertical polarization states of a single photon, respectively. For the $\sigma_x$ Hamiltonian the excited state is $| e \rangle = (|H\rangle+|V\rangle)/\sqrt{2}$ and the ground state is $| g \rangle= (|H\rangle-|V\rangle)/\sqrt{2}$.


\item For the $\sigma_z$ Hamiltonian, the polarization state of the photon is mapped to the logical state of the system as $\ket 0 = \ket
 H$ and $\ket 1 = \ket V$. The two different path of the interferometer
 serves as the logical qubit of the ancilla. The system is prepared in  the state $\alpha|0\rangle+\beta|1\rangle$ (blue basis, see Fig~\ref{fig:setup}d) and the ancilla qubit is prepared as $|0\rangle$ (red basis) (step 1).
 
\item In order to simulate cooling with the Hamiltonian $\sigma_{x}$, the $\sigma_{x}$ transform was applied to
the input and output ports: we rotate the basis from $|H\rangle$ and $|V\rangle$ to $1/\sqrt{2}(|H\rangle+|V\rangle)$ and $1/\sqrt{2}(|H\rangle-|V\rangle)$ This transform consists of a half-wave plate (HWP) with the angle setting at $22.5^{\circ}$ and a tiltable quarter-wave plate (QWP). The logical circuit corresponding to the experimental setup is shown in Fig.~\ref{fig:setup}(c), and the corresponding sequence of states is shown in Fig.~\ref{fig:setup}(d).

\item{The input photon pulses for the simulation were generated by a laser
with a central wavelength of 800 nm and a repetition rate of 76
MHz, attenuated to the single photon level.}

\item{The measurement bases for the tomography were set by a QWP, a HWP and a polarization beam splitter (PBS). The photon was finally detected by a single photon detector (SPD) equipped with a 3 nm interference
filter (IF). The probability of each output can be directly deduced from the single photon counts of the tomography measurements.}

\end{enumerate}

\subsection{Step-by-step instruction of the implementation of the cooling module}

\begin{enumerate}
\item In our setup, a photon is generated from the single photon source shown in Fig.~\ref{fig:q_circuit2exp}c.

\item The photon is then passed through the state preparation module consisting of a polarization beam splitter (PBS), half-wave plate (HWP) and a quarter-wave plate (QWP). The resulting state is denoted as  $\alpha|0\rangle+\beta|1\rangle$  (blue basis, see Fig~\ref{fig:q_circuit2exp}d). The state associated with the path degrees of freedom is denoted as $|0\rangle$. Combining the two, we have the state 
\begin{equation}
\left( {\alpha \left| 0 \right\rangle  + \beta \left| 1 \right\rangle } \right)\left| 0 \right\rangle \quad
\end{equation}
as the input state for the cooling module. This corresponds to step 1 of Fig.~\ref{fig:q_circuit2exp}.

\item{The incident photon is then split by a polarization beam splitter, the horizontal polarized photons propagate directly while the vertical polarized photons reflect. This can be considered as a controlled-NOT operation. The resulting state becomes 
\begin{equation}
\alpha \left| 0 \right\rangle \left| 0 \right\rangle  + \beta \left| 1 \right\rangle \left| 1 \right\rangle \quad.
\end{equation}
This corresponds to step 2 of Fig.~\ref{fig:q_circuit2exp}}.

\item{For each path, a unitary operation is realized by using two half-wave plates (HWP), and a quartz plate compensator (PC). The resulting state is 
\begin{equation}
\alpha \left( {a\left| 0 \right\rangle  + b\left| 1 \right\rangle } \right)\left| 0 \right\rangle  + \beta \left( {a\left| 0 \right\rangle  + b\left| 1 \right\rangle } \right)\left| 1 \right\rangle \quad.
\end{equation}
This corresponds to step 3 and step 4 in Fig.~\ref{fig:q_circuit2exp}. }
\item{The two paths are then recombined on the same polarization beam splitter (PBS). The same controlled-NOT is applied, and the resulting is 
\begin{equation}
\left( {\alpha a\left| 0 \right\rangle  + \beta b\left| 1 \right\rangle } \right)\left| 0 \right\rangle  + \left( {\alpha b\left| 1 \right\rangle  + \beta a\left| 0 \right\rangle } \right)\left| 1 \right\rangle \quad.
\end{equation}
This corresponds to step 5 in Fig.~\ref{fig:q_circuit2exp}.}

\item{For the path $\left| 1 \right\rangle$, a half-wave plate (controlled-not gate) is applied to the polarization state, giving the final state
\begin{equation}
(\alpha a|0\rangle+\beta b|1\rangle)|0\rangle+(\alpha b|0\rangle+\beta a|1\rangle)|1\rangle \quad.
\end{equation}
Note that the unitaries are $U_{0}|0\rangle=U_{1}|1\rangle=a|0\rangle+b|1\rangle$, where
 $a=(1+ie^{i\theta})/2$ and $b=(1-ie^{i\theta})/2$. This corresponds to step 6 in Fig.~\ref{fig:q_circuit2exp}.
}

\item{At one output port, denoted as 0, the photon state is cooled and was sent to the next cooling module. At the other output port, denoted by 1, the photon is heated and was either rejected or recycled (re-prepared to the initial state and sent to the next cooling module).}

\end{enumerate}

\section{4. Additional experimental results}

\begin{figure*}[b]
\begin{center}
\includegraphics [width= 0.8 \columnwidth]{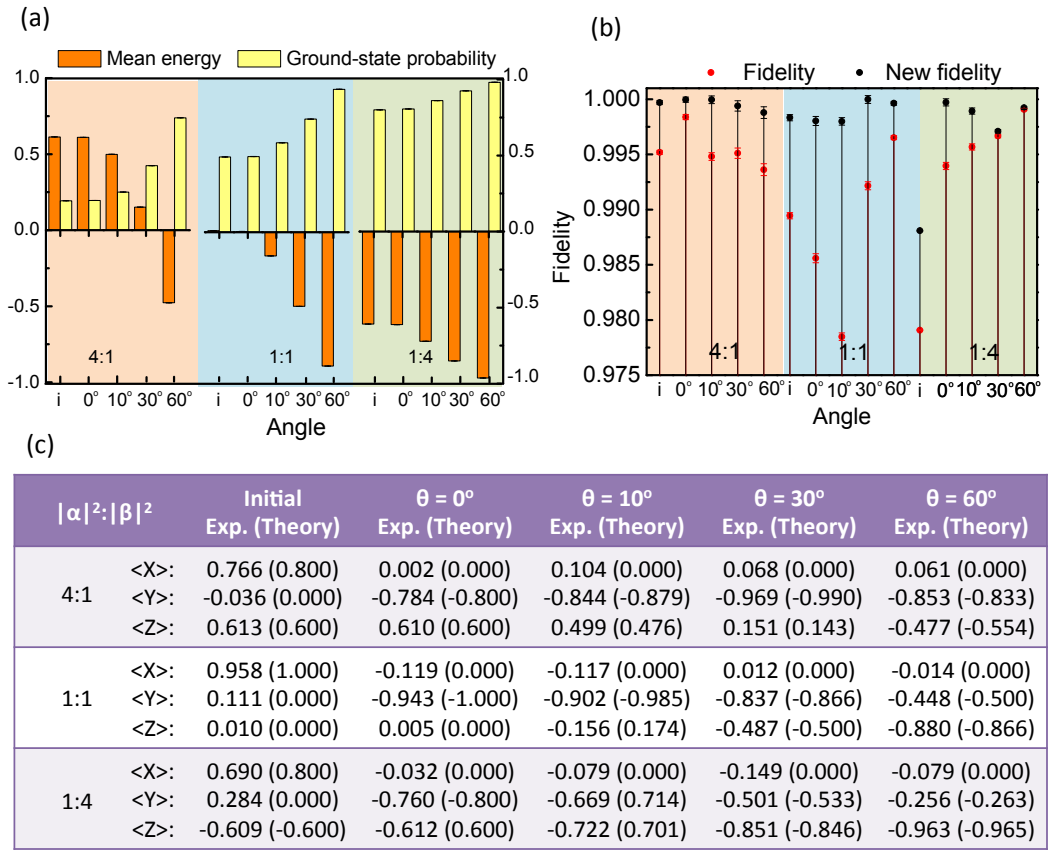}
\end{center}
\caption{(Color online). Experimental results for a single cooling
  module with initial state $(|e\rangle+|g\rangle) /\sqrt{2}$ as a
  function of the energy bias angle $\theta$.  (a) Mean energies and ground-state
  probabilities of the reconstructed states for different energy bias angles. Different color panes
  represent the cases with different initial input states (``i" $=\alpha \ket e + \beta \ket g$) with relative populations
  $|\alpha|^{2}:|\beta|^{2}=4\!:\!1$, $1\!:\!1$, $1\!:\!4$. (b) The fidelities of the
  reconstructed states. Red squares represent the experimental
  fidelities. Black dots represent revised fidelities. Error bars are
  the corresponding standard deviations. (c) Tomography data.}
\label{fig:1stcycle}
\end{figure*}

Figure~\ref{fig:1stcycle} shows the experimental results for the state
transformation carried out by a single cooling module which simulates
the Hamiltonian
\begin{equation}
H_s=\sigma_{z} \;.
\end{equation}
The output state (photon) at the output port $1$ (which corresponds to heating) is rejected. The ratio
between $|e\rangle$ and $|g\rangle$ of the output state decreases by a factor
\begin{equation}
(1-\sin\theta)/(1+\sin\theta) \;.
\end{equation}
As a result, the output state becomes closer to the ground state with increasing $\theta$, which is verified
by our experimental results. For an energy bias angle
$\theta=0^{\circ}$ the state simply rotates in the equatorial plane of
the Bloch sphere.  For the input state
\begin{equation}
(|e\rangle+|g\rangle)/\sqrt{2} \;,
\end{equation}
the output state as a function of the energy bias angle $\theta$ has Bloch sphere angles
\begin{equation}
\langle
\sigma_y \rangle = \cos \theta \quad {\rm and} \quad \langle \sigma_z \rangle = -\sin
\theta \;.
\end{equation}

Figure~\ref{fig:1stcycle}a shows the mean energies and ground state probabilities for different
initial states (organized in panels with different colors) and
different energy bias angles. The fidelities of the reconstructed
states are shown in Fig.~\ref{fig:1stcycle}b. Red squares represent
the experimental results and black dots represent the revised
fidelities (see Methods Summary). The fidelities of the final states
are larger than 0.978. Due to the compensation of symmetry noise, the
revised fidelities are larger.

\begin{figure}[t]
\begin{center}
\includegraphics [width= 1 \columnwidth]{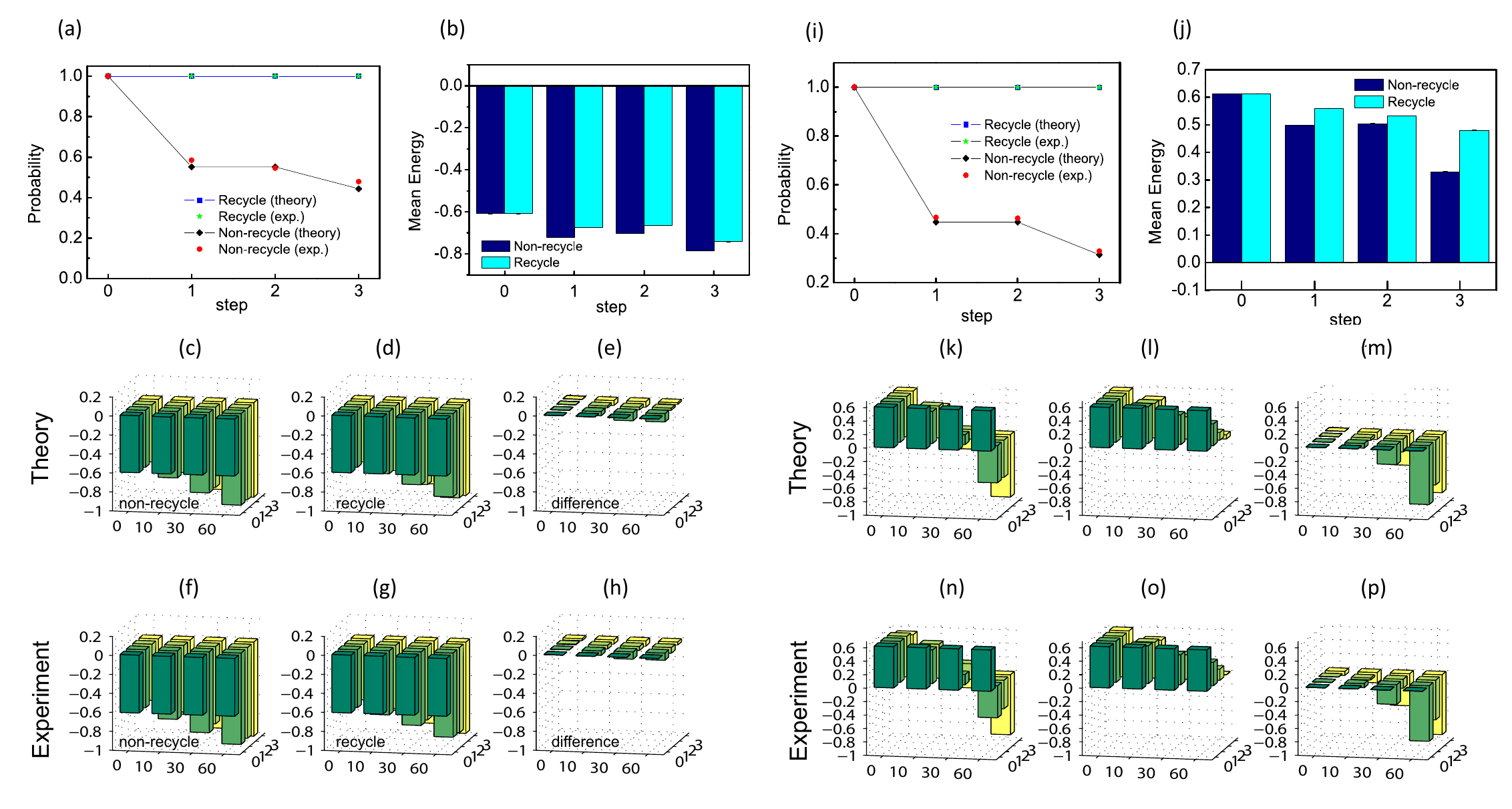}
\end{center}
\caption{(Color online). Comparison of simulated cooling with evaporative
  and recycling strategies for the input states
  $(1/\sqrt{5})|e\rangle+(2/\sqrt{5})|g\rangle$ (left) and
  $(2/\sqrt{5})|e\rangle+(1/\sqrt{5})|g\rangle$ (right) for Hamiltonian $\sigma_z$. (a) Experimental
  results for the success probabilities in each cooling step with evaporative and
  recycling strategies for input state
  $(1/\sqrt{5})|e\rangle+(2/\sqrt{5})|g\rangle$ (and
  $\theta=10^{\circ}$). (b) Experimental results for the corresponding
  mean energies. (c-e) theoretical results for the mean energies for the evaporative strategy,
the recycling strategy, and their difference. (f-h) Corresponding experimental results. Figures (i-p) show the same data but now with input state $(2/\sqrt{5})|e\rangle+(1/\sqrt{5})|g\rangle$.  } \label{fig:S1}
\end{figure}

Figure~\ref{fig:S1} shows the experimental results with evaporative and recycling strategies with initial input states
\begin{equation}
\left( {1/\sqrt 5 } \right)\left| e \right\rangle  + \left( {2/\sqrt 5 } \right)\left| g \right\rangle \quad {\rm and} \quad
\left( {2/\sqrt 5 } \right)\left| e \right\rangle  + \left( {1/\sqrt 5 } \right)\left| g \right\rangle
\end{equation}
respectively for the left and right panel, for Hamiltonian $\sigma_{z}$ (see also Fig.~\ref{fig:fullcycle}(f-m)). Figure~\ref{fig:S1}(a) depicts the success probabilities at each step with evaporative and recycling strategies for input state $(1/\sqrt{5})|e\rangle+(2/\sqrt{5})|g\rangle$ ($\theta=10^{\circ}$).  Figure~\ref{fig:S1}(b) shows the mean energies with evaporative and recycling strategies. Figures~\ref{fig:S1}(c-e) indicate the
theoretical predictions for the mean energies for the cases with
evaporative, with recycling, and their
difference. Figures~~\ref{fig:S1}(f-h) are the corresponding
experimental results. The difference between the cases with evaporative and
recycling is small compared to the case of
Fig.~\ref{fig:fullcycle}.  Figure~\ref{fig:S1}(i-p) shows the exact
same plots for the input state
$(2/\sqrt{5})|e\rangle+(1/\sqrt{5})|g\rangle$. The difference between the
cases with evaporative and recycling is larger than that in
Fig.~\ref{fig:S1}(c-h), as shown in Fig.~\ref{fig:S1}(k-p).

\begin{figure}[h]
\begin{center}
\includegraphics [width= 0.6 \columnwidth]{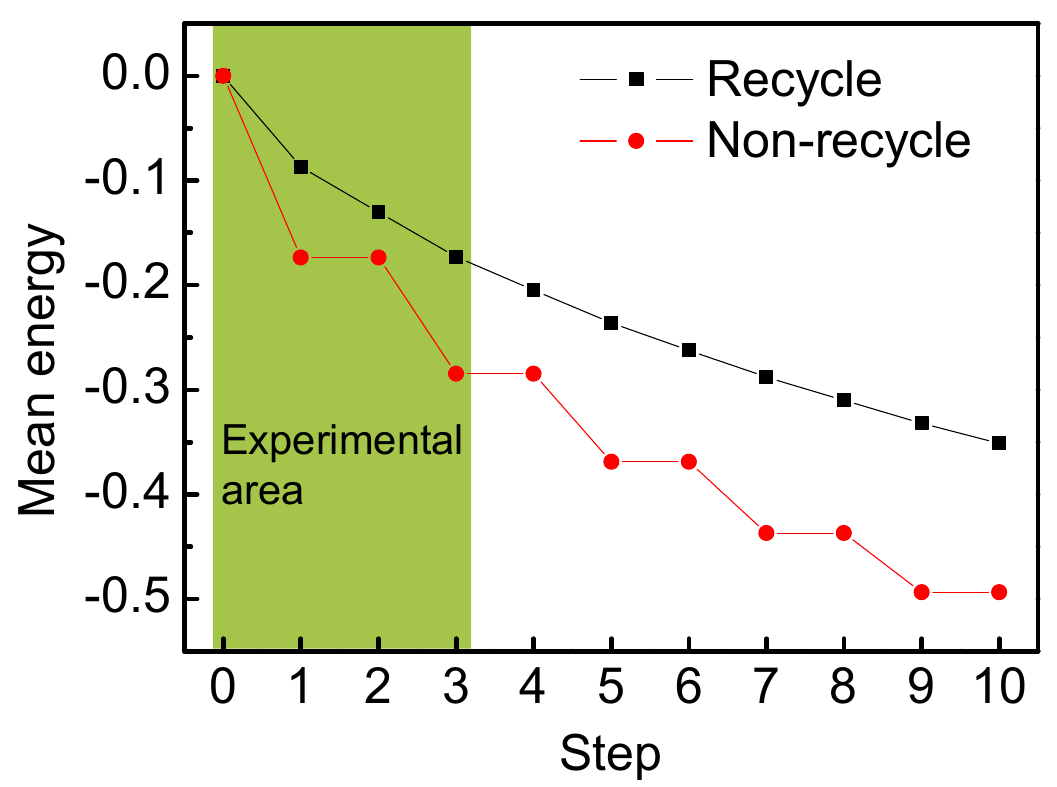}
\end{center}
\caption{(Color online). Theoretical results for the mean energies
  with ten cooling steps and Hamiltonian $\sigma_{z}$ . Black squares
  represent the results with recycling and the red dots represent the
  case with evaporative. The initial state is
  $(|e\rangle+|g\rangle) /\sqrt{2}$ and the energy bias angle $\theta$
  is set to $10^{\circ}$.} \label{fig:S4}
\end{figure}

In the experiment, we performed three steps of cooling, we show the theoretical calculation for up to ten cooling steps in Fig.~\ref{fig:S4}.  The mean energies with evaporative are always smaller than the case with recycling.